\documentclass{article}
\usepackage{spconf,amsmath,graphicx}

\usepackage{hyperref}       
\usepackage{url}            
\usepackage{booktabs}       
\usepackage{amsfonts}       
\usepackage{nicefrac}       
\usepackage{microtype}      
\usepackage{xcolor}         

\usepackage{graphicx}
\usepackage{amsmath}
\hypersetup{colorlinks=true,linkcolor=blue, linktocpage}

\title{DATAR: DEFORMABLE AUDIO TRANSFORMER FOR AUDIO EVENT RECOGNITION}

\name{Wentao Zhu} 
\begin{document}
%
\maketitle
\begin{abstract}
Transformers have achieved promising results on a variety of tasks. However, the quadratic complexity in self-attention computation has limited the applications, especially in low-resource settings and mobile or edge devices. Existing works have proposed to exploit hand-crafted attention patterns to reduce computation complexity. However, such hand-crafted patterns are data-agnostic and may not be optimal. Hence, it is likely that relevant keys or values are being reduced, while less important ones are still preserved. Based on this key insight, we propose a novel deformable audio Transformer for audio recognition, named DATAR, where a deformable attention equipping with a pyramid transformer backbone is constructed and learnable. Such an architecture has been proven effective in prediction tasks,~\textit{e.g.}, event classification. Moreover, we identify that the deformable attention map computation may over-simplify the input feature, which can be further enhanced. Hence, we introduce a learnable input adaptor to alleviate this issue, and DATAR achieves state-of-the-art performance. 
\end{abstract}
\begin{keywords}
Deformable Audio Transformer, Audio Adaptor, Event Classification
\end{keywords}
\section{Introduction}
Transformer~\cite{vaswani2017attention} is firstly introduced to address natural language processing tasks,~\textit{e.g.}, machine translation. Recently, it has shown great potential in the field of audio and speech recognition~\cite{gulati2020conformer,kong2020sound,kazakos2021slow}. Transformer-based models,~\textit{i.e.}, self-attention, are good at modeling long-range dependencies, which are proven to achieve superior performance in the regime of a large amount of learnable model parameters and training data. However, the quadratic computational complexity of the Transformer to the number of input tokens limits the applications, especially in low-resource settings. Specifically, the superfluous number of keys to attend per query yields high computational costs and slow convergence.

\begin{figure}[ht!]
  \centering
  \includegraphics[width=0.95\linewidth,trim=5 4 4 9,clip]{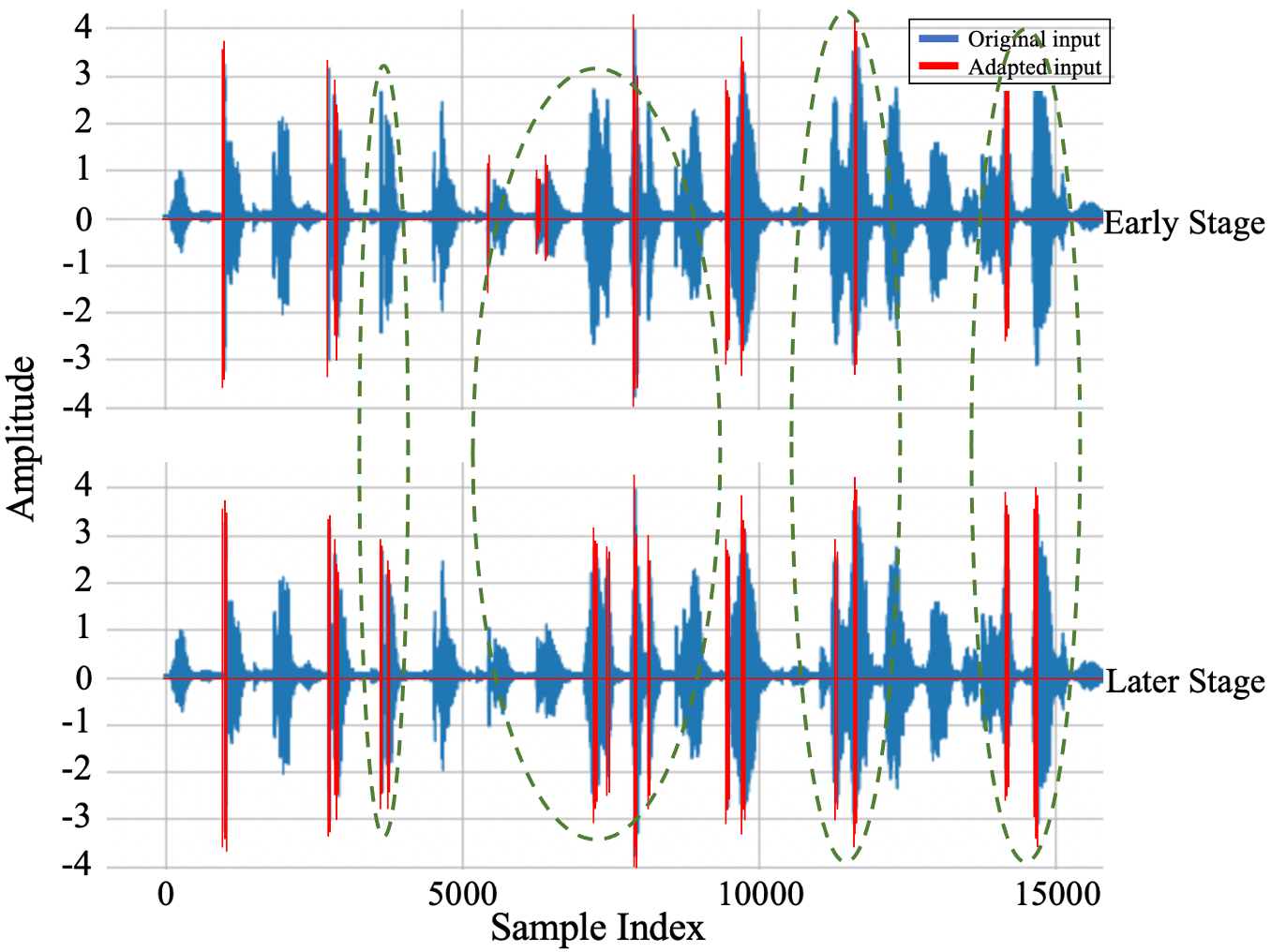}
  \caption{Visualization of the effectiveness of proposed learnable input adaptor, which enhances relatively important parts of the original signal, a log-compressed mel-scaled spectrogram, to become more distinguishable, colored by green dash ovals.}
  \label{figure1}
\end{figure}
To address the issue of excessive attention computation, \cite{zaheer2020big,beltagy2020longformer,liu2021swin,wang2021pyramid} have leveraged carefully designed efficient attention patterns to alleviate the computation complexity. However, since the hand-crafted attention patterns are data-agnostic and may not be optimal, it probably leads to that relevant keys or values being dropped, while less important ones are still preserved. To tackle the issue of hand-crafted attention patterns, \cite{dai2017deformable} exploits an efficient deformable attention to generate the candidate key or value set for a given query flexibly. Moreover, the mechanism makes the model be able to adapt to each individual input. Recently, deformable attention-based networks have yielded promising results on many challenging tasks~\cite{zhu2020deformable,wang2022deformable,xia2022vision}. Hence, this motivates us to explore the deformable attention mechanism in audio transformers.

In this work, a deformable audio transformer, DATAR, is proposed in Fig.~\ref{figure2}. DATAR is equipped with a powerful pyramid backbone based on the deformable attention. Such a backbone is popular in prediction tasks,~\textit{e.g.}, image classification~\cite{xia2022vision}. Based on the observation in~\cite{cao2019gcnet,zhou2021deepvit}, global attention typically results in almost the same attention patterns for various queries. Hence, a trainable audio offset generator (AOG) is introduced to learn a few groups of query-agnostic offsets to shift keys and values to important regions. Such a design not only helps hold a linear space complexity, but also introduces a deformable attention pattern to transformer backbones. Specifically, for each self-attention module, reference points are first generated as uniform grids, which are the same across the input data. Then, the introduced AOG takes input as the query features and generates the corresponding offsets for all the reference points. Hence, the candidates of keys or values are shifted towards important regions. This augments the self-attention module with higher efficiency and flexibility to capture more informative features.

Furthermore, we identify that the training of DATAR faces an accuracy improvement bottleneck, which is most likely due to that the deformable attention map calculation over-simplifies the input feature, leading to information loss~\cite{xia2022vision}. We address this challenge by introducing a learnable input adaptor that adds learnable signals to the original input for more accurate deformable audio attention map computation. As these learned signals make the important parts of the original input become more distinguishable, the effectiveness of deformable audio attention is improved in Fig.~\ref{figure1}.


\section{Related work}
Transformers~\cite{vaswani2017attention} have been validated in the fields of natural language processing and computer vision. Recently, \cite{gong2021ast,chen2022hts,kazakos2021slow} introduce the Transformer into audio and speech processing and achieve the state-of-the-art performance. AST~\cite{gong2021ast} proposes a convolution-free audio spectrogram Transformer, which is directly applied to an audio spectrogram and capable of capturing long-range global context even in the lowest layers. However, it requires large GPU memory and long training time, which limits the model’s scalability in audio related tasks~\cite{chen2022hts}. MAST and HTS-AT~\cite{zhu2023multiscale,chen2022hts,zhu2022multiscale,zhu2022selective} introduce a hierarchical structure into the AST for audio event detection. Inspired by the two pathways in the human auditory system, \cite{kazakos2021slow} introduces a two-stream ConvNet for audio recognition,~\textit{i.e.}, fusing slow and fast streams with multi-level lateral connections. The slow pathway has a higher channel capacity, while the fast pathway has fewer channels and operates at a fine-grained temporal resolution. Different from the AST, \cite{miyazaki2020convolution,kong2020sound,gulati2020conformer} combine the Transformer architecture and ConvNet for audio processing. \cite{miyazaki2020convolution,kong2020sound} stack a Transformer on top of a ConvNet. \cite{gulati2020conformer} combines a Transformer and a ConvNet in each model block. Other efforts combine ConvNets with simpler attention modules~\cite{kong2020panns,gong2021psla,rybakov2020streaming}. The main model structure of DATAR is convolution-free.  

The main challenge of self-attention is the quadratic computation complexity~\cite{vaswani2017attention,dosovitskiy2020image}. \cite{chen2021regionvit,dong2022cswin,liu2021swin,pan2022integration,wang2021pyramid,yang2021focal,zhang2021multi} have proposed various efficient methods to address this issue. The improvements focus on learning multiscale features for prediction tasks or efficient attention mechanisms. These efficient attention mechanisms include global tokens~\cite{chen2021regionvit,jaegle2021perceiver,bai2021visual}, windowed attention~\cite{dong2022cswin,liu2021swin}, dynamic token sizes~\cite{wang2021not}, and focal attention~\cite{yang2021focal}.
Deformable convolution~\cite{dai2017deformable,zhu2019deformable} is a powerful technique to attend to flexible locations conditioned on input data. Recently, \cite{zhu2020deformable,yue2021vision,chen2021dpt} have applied it to vision Transformers (ViT). Deformable DETR~\cite{zhu2020deformable} improves the convergence speed of DETR~\cite{carion2020end} by selecting a few keys for each query on the top of a ConvNet backbone. \cite{yue2021vision,chen2021dpt} introduce deformable modules to refine visual tokens. \cite{yue2021vision} introduces a spatial sampling module before a ViT backbone to improve visual tokens. \cite{chen2021dpt} proposes to refine patches across stages by deformable patch embeddings. We are the first to exploit deformable attention in audio event classification.

\section{Methodology}
The main idea of deformable attention is to flexibly generate discriminative key and value points for each query token, compared with a regular grid sampling strategy in a conventional Transformer. This learned deformable sampling strategy yields a more useful and discriminative dot-product matrix and output. To ease the computational complexity of deformation generation, we leverage the redundancy in the signal, and the offset network is conducted in 4$\times$ subsampled tokens by a convolutional neural network followed by bilinear interpolation.
\subsection{Deformable Audio Transformer}
Let $X \in \mathbb{R}^{h\times T}$ be an audio spectrogram as input for DATAR, where $h$ is the number of triangular mel-frequency bins, and $T$ is the temporal length. After the patch embedding, which can be a convolutional block conducted in the audio spectrogram, we obtain the embedding token matrix $A \in \mathbb{R}^{N\times C}$, where $C$ is the embedding dimension and $N$ is the number of tokens. 
A multi-head self-attention (MHSA) block with $M$ heads is formulated:
\begin{equation}
    \begin{aligned}
  q &= A W_{q}, \quad
  k = A W_{k}, \quad
  v = A W_{v}, \nonumber \\
  z^{(m)} &= \sigma(q^{(m)}k^{(m)\top}/\sqrt{d})v^{(m)}, \, m=1, \cdots,M, \nonumber \\ 
  z &=\textup{Concat}\left( z^{(1)}, \cdots,z^{(M)} \right)W_{o},
\end{aligned}\label{eq:1}
\end{equation}
where $\sigma(\cdot)$ denotes the softmax function, and $d = C/M$ is the dimension of each head. $z^{(m)}$ denotes the embedding output from the $m$-th attention head, $q^{(m)}$, $k^{(m)}$, $v^{(m)}$ $\in \mathbb{R}^{N \times d}$ denote query, key, and value embeddings respectively. $W_{q}$, $W_{k}$, $W_{v}$, $W_{o}$ $\in \mathbb{R}^{C \times C}$ are the projection matrices. To build up a Transformer block, an MLP block with two linear layers and a GELU activation is usually adopted to provide nonlinearity.
With layer normalization~\cite{ba2016layer} (LN) and identity shortcuts, the $l$-th Transformer block is formulated as:
\begin{align}
  z^{\prime}_{l} &= \textup{MHSA}(\textup{LN}(z_{l-1}))+z_{l-1}, \nonumber \\
  z_{l} &= \textup{MLP}(\textup{LN}(z^{\prime}_{l}))+z^{\prime}_{l}.
\end{align}

\begin{figure}[t!]
  \centering
  \includegraphics[width=0.95\linewidth,trim=5 4 4 9,clip]{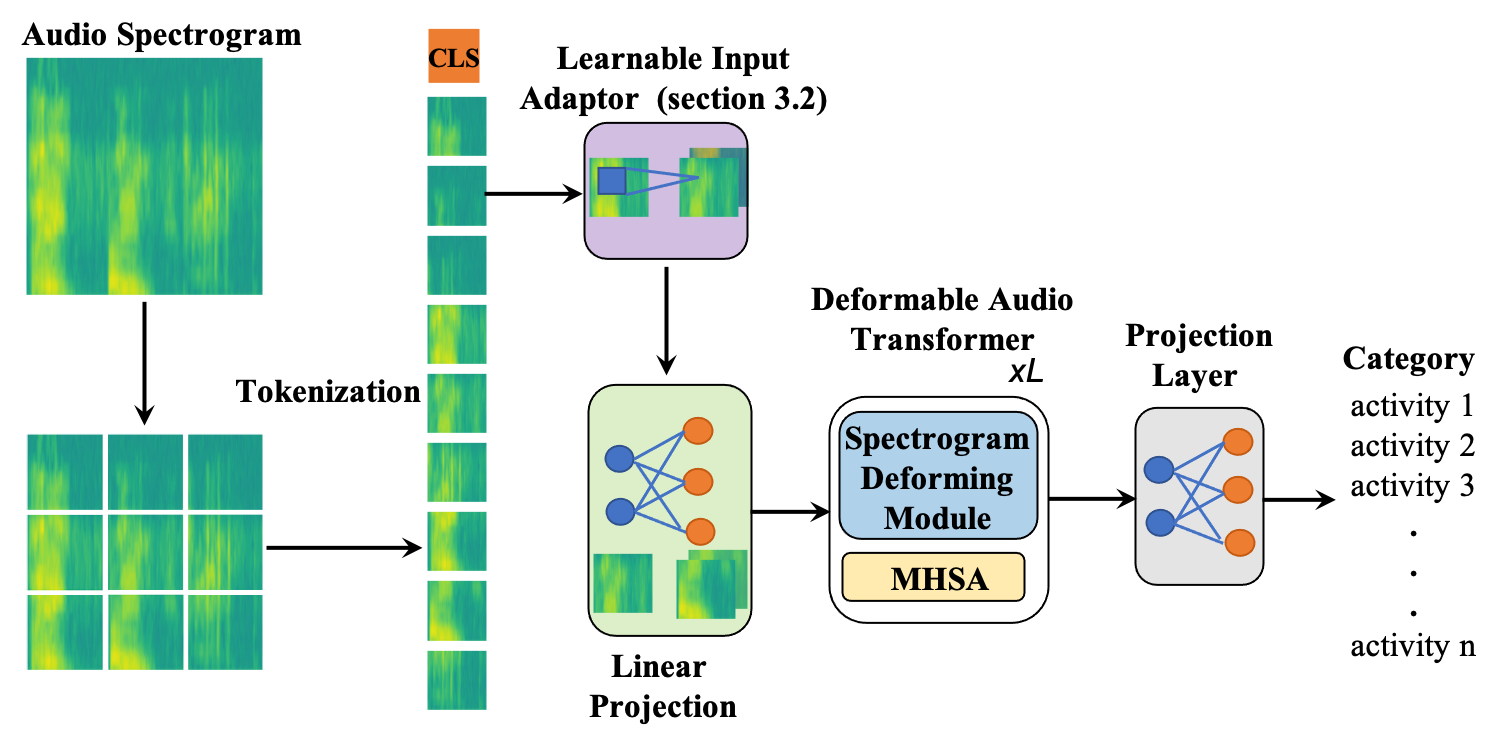}
  \caption{Illustration of deformable audio Transformer.}
  \label{figure2}
\end{figure}

\begin{figure}[ht!]
  \centering
  \includegraphics[width=0.95\linewidth]{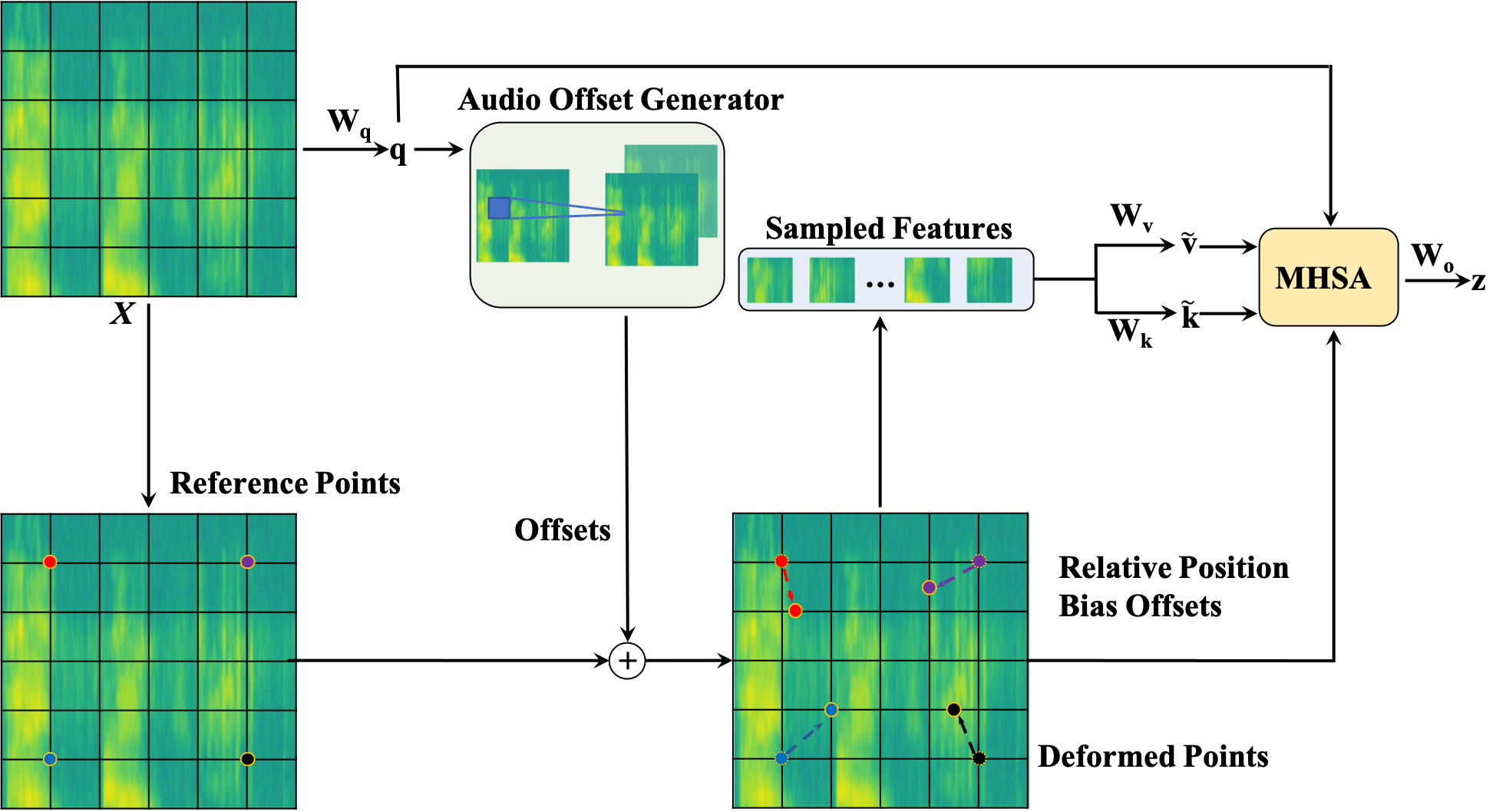}
  \caption{Illustration of spectrogram deforming processing. Audio offset generator in $\S3.1$ calculates offsets used in deformable attention. When applying the deformable attention on the audio spectrum, for each query token, we select the most informative spectrum patch to conduct the self-attention. Based on the query location, DATAR learns the offset to localize the target spectrum patches.}  
  \label{figure3}
\end{figure}

\noindent\textbf{Deformable attention module}
Given the input feature map $X \in \mathbb{R}^{h \times T}$, we generate a uniform grid of points $p \in \mathbb{R}^{h_{G} \times T_{G} \times 2}$ as the references in Fig.~\ref{figure2}. Specifically, the grid size is down-sampled from the input feature map size by a factor $r$, $h_{G} = h/r$, $T_{G} = T/r$. The values of reference points are linearly spaced 2D coordinates $\{(0, 0), \cdots ,(h_{G} - 1, T_{G} - 1)\}$. Then we normalize them to the range $[-1, 1]$ according to the grid shape $h_{G} \times T_{G}$, in which $(-1, -1)$ indicates the top-left corner and $(1, 1)$ indicates the bottom-right corner.

Without loss of generality, we still use $X$ as input for each block and omit the patch embedding and reshaping operators. To obtain the offset for each reference point, the feature maps are projected linearly to the query tokens $q=X W_{q}$, and then fed into a lightweight subnetwork $\theta_\textup{offset}(\cdot)$ to generate the offsets $\Delta p = \theta_\textup{offset}(q)$. To stabilize the training process, we scale the amplitude of $\Delta p$ by a predefined factor $s$ to prevent too large offset,~\textit{i.e.}, $\Delta p \gets  s ~ \textup{tanh} (\Delta p)$. Then the features are sampled at the locations of deformed points as keys and values, followed by projection matrices:
\begin{align}
  q = X W_{q}, \quad
  \tilde{k} = \tilde{X}W_{k}, \quad
  \tilde{v} = \tilde{X}W_{v}, \nonumber \\
   \textup{with} ~ \Delta p = \theta_\textup{offset}(q), \quad \tilde{X} = \phi(X; p+\Delta p),
\end{align}
where $\tilde{k}$ and $\tilde{v}$ represent the deformed key and value embeddings respectively. Specifically, we set the sampling function $\phi(\cdot;\cdot)$ to a bilinear interpolation to make it differentiable:
\begin{equation}
    \begin{aligned}
\phi(z;(p_{x},p_{y}))=\sum_{(r_{x},r_{y})}g(p_{x},r_{x})g(p_{y},r_{y})z[r_{y},r_{x},:],
\end{aligned}\label{eq:4}
\end{equation}
where $g(a,b)=\textup{max}(0,1- \left| a-b \right|)$ and $(r_{x},r_{y})$ indexes all the locations on $z \in \mathbb{R}^{h \times T \times C}$. As $g$ would be non-zero only on the four integral points closest to $(p_{x}, p_{y})$, it simplifies Eq.~\eqref{eq:4} to a weighted average of the four locations. We perform multi-head attention on $q$, $\tilde{k}$, $\tilde{v}$ and adopt relative position offsets $R$. The output of an attention head is formulated as:
\begin{align}
z^{(m)}=\sigma \left( {q^{(m)}\tilde{k}^{(m)\top} / \sqrt{d} + \phi(\hat{B};R)} \right) \tilde{v}^{(m)},
\end{align}
where $\phi(\hat{B};R) \in \mathbb{R}^{hT \times h_{G}T_{G}}$ corresponds to the position embedding following previous work~\cite{liu2021swin}.  

\noindent\textbf{Audio offset generator}
A subnetwork is adopted for offset generation, which consumes the query features and outputs the offset values for reference points. Considering that each reference point covers a local $s \times s$ region where $s$ is the largest value for offset, the generation network should also have the perception of the local features to learn reasonable offsets. Therefore, we implement the subnetwork as two convolution modules with a nonlinear activation in Fig.~\ref{figure3}. The input features are first passed through a $5 \times 5$ depth-wise convolution to capture local features. Then, a GELU activation and a $1 \times 1$ convolution is adopted to get the 2D offsets. It is also worth noticing that the bias in $1 \times 1$ convolution is dropped to alleviate the compulsive shift for all locations.


\noindent\textbf{Deformable relative position bias}
Relative position bias encodes the relative positions between every pair of query and key, which augments the vanilla attention with spatial information. Considering a feature map with shape $h \times T$, its relative coordinate displacements lie in the range of $\left[ -h, h \right]$ and $\left[ -T, T \right]$ at two dimensions, respectively. In Swin Transformer~\cite{liu2021swin}, a relative position bias table $\hat{B} \in \mathbb{R}^{(2h-1) \times (2T-1)}$ is constructed to obtain the relative position bias by indexing the table with the relative displacements in two directions. Since our deformable attention has continuous positions of keys, we compute the relative displacements in the normalized range $\left[ -1, 1 \right]$, and then interpolate $\phi(\hat{B}; R)$ in the parameterized bias table $\hat{B} \in \mathbb{R}^{(2h-1) \times (2T-1)}$ by the continuous relative displacements in order to cover all possible offset values.

\subsection{Learnable Input Adaptor}
To further increase the accuracy, we apply a 2D convolution over an input signal composed of several input planes. In the simplest case, the output value of the layer with input size $(C, h, T)$ and output can be defined as:
\begin{align}
 \textup{out}(C_{j}) =\textup{input}(C_{j}) + \lambda (\textup{b}(C_{j}) +\sum_{k=0}^{C-1}\textup{W}(C_{j}, k) \ast \textup{input}(k)),
\end{align}
where $C_j$ is the channel index, $\lambda$ is used to tune the strength of adaptor, $\textup{b}$ and $\textup{W}$ are learnable parameters to enhance input signals, $\ast $ is a 2D cross-correlation operator, $C$ denotes the number of channels, $h$ is the number of triangular mel-frequency bins, and $T$ is the temporal length, $\textup{input}(k)$ denotes the $k$-th channel of $\textup{input}$.  

\section{Experiments}
We experiment with three audio event classification datasets – Kinetics-Sounds~\cite{arandjelovic2017look,kay2017kinetics}, Epic-Kitchens-100~\cite{Damen2021RESCALING,Damen2018EPICKITCHENS,Damen2021PAMI}, and VGGSound~\cite{chen2020vggsound}. The input is TorchAudio-based fbank, which is a log-compressed mel-scaled spectrogram and the same as AST. The frame length is 1,024 and the number of triangular mel-frequency bins is 128, which are the same as AST. The sampling frequency is 43,000 to cover 10 seconds of audio. We use cross-entropy loss. On Epic-Kitchens-100, we employ two classification heads for verb and noun.


Kinetics-Sounds~\cite{kay2017kinetics,arandjelovic2017look} is a commonly used subset of the Kinetics-400 dataset~\cite{kay2017kinetics}, which is composed of 10-second audio data from YouTube. The dataset collection protocol described in~\cite{xiao2020audiovisual} is followed. $22,914$ valid training audio data and $1,585$ valid test audio data are collected.  %
Epic-Kitchens-100~\cite{Damen2021PAMI} consists of 90,000 variable length egocentric clips spanning 100 hours capturing daily kitchen activities. The dataset formulates each action into a verb and a noun. There are 67,217 training and 9,668 test samples after we remove categories with no training sample, which results in 97 verbs and 293 nouns. 
VGGSound~\cite{chen2020vggsound} consists of about $200,000$ 10-second video clips and $309$ categories ranging from human actions and sound-emitting objects to human-object interactions. After removing invalid audio data, $159,223$ valid training audio data and $12,790$ valid test audio data are collected.  

\noindent\textbf{Hyperparameters} In all the experiments, we follow AST [21] setting in which model is pre-trained on ImageNet-1k. For hyperparameters in DATAR, we follow DAT~\cite{xia2022vision} and use ImageNet-1K publicly available pretrained weights. AdamW is used with the learning rate of 0.00001. The numbers of epochs are set as 300, 100, and 50 for Kinetics-Sounds, Epic-Kitchens-100, and VGGSound. We employ the code of AST to obtain the results on these datasets.  

\noindent\textbf{Comparison with state-of-the-arts}
According to Table~\ref{table:1},~\ref{table:2}, and~\ref{table:3}, the comparative results show that the proposed method is effective, and it outperforms the previous state-of-the-arts by a large margin. Specifically, the DATAR outperforms the previous best methods by 18.9\%, 3.2\% and 0.9\% in terms of top-1 accuracy on the three datasets. The main reason is that, the introduced deformable audio attention and the learnable input adaptor help the model effectively capture the informative audio signals which are used to boost the model performance. For computation complexity, our work achieves better accuracy and consumes 93.3G MACs, while AST uses 103.4G MACs on VGGSound.

\noindent\textbf{Ablation study}
To validate the effectiveness of the proposed deformable audio attention and the input adaptor, we conduct the ablation study of `without deformable', `with deformable', adding Gaussian perturbation to the input instead of learnable adaptor, adding Laplacian perturbation, different strengths, 0.2, 0.005, of $\lambda$ in Table~\ref{table:4}. Deformable audio attention helps improve the model performance,~\textit{i.e.}, $1.1\%$ improvement in terms of top-1 accuracy on Kinetics-Sounds dataset. Based on Table~\ref{table:4}, the proposed input adaptor is effective and improves the baseline model (DATAR with deformable) by $1\%$ in top-1 accuracy. Compared with Gaussian and Laplacian perturbations, the proposed DATAR with the learnable input adaptor surpasses the two perturbation based methods by 1.3\% and 1\%. This demonstrates the effectiveness of learnable input adaptor.

\begin{table}[t!]
\small
\centering
\scalebox{1.0}{
\begin{tabular}{c|ccc}
\toprule
\multicolumn{2}{c|}{\textbf{Model}}               & \textbf{Top-1 acc} & \textbf{Top-5 acc}  \\ 
\midrule
\multicolumn{2}{c|}{AST~\cite{nagrani2021attention}}    & 52.6	& 71.5  \\ 
\midrule
\multicolumn{2}{c|}{DATAR (deformable+adaptor)}    & \textbf{71.5}	& \textbf{90.5}	 \\ 
\bottomrule
\end{tabular}}
\caption{Comparison with state-of-the-arts on Kinetics-Sounds.}  
\label{table:1}
\end{table}

\begin{table}[t!]
\small
\centering
\scalebox{1.0}{
\begin{tabular}{c|ccc}
\toprule
\multicolumn{2}{c|}{\textbf{Model}}               & \textbf{Top-1 acc} & \textbf{Top-5 acc}  \\ 
\midrule
\multicolumn{2}{c|}{AST~\cite{nagrani2021attention}}    & 52.3	& 78.1  \\ 
\midrule
\multicolumn{2}{c|}{Chen et al.~\cite{chen2020vggsound}}    & 48.8	& 76.5  \\ 
\midrule
\multicolumn{2}{c|}{AudioSlowFast~\cite{kazakos2021slow}}    & 50.1	& 77.9	 \\ 
\midrule
\multicolumn{2}{c|}{DATAR (deformable+adaptor)}    & \textbf{55.5}	& \textbf{80.2}	 \\ 
\bottomrule
\end{tabular}}
\caption{Comparison with state-of-the-arts on VGGSound.}  
\label{table:2}
\end{table}

\begin{table}[t!]
\small
\centering
\scalebox{1.0}{
\begin{tabular}{c|cccc}
\toprule
\multicolumn{2}{c|}{\textbf{Model}}               & \textbf{Verb} & \textbf{Noun} & \textbf{Action}  \\ 
\midrule
\multicolumn{2}{c|}{AST~\cite{nagrani2021attention}}    & 44.3	& 22.4 & 13.0 \\ 
\midrule
\multicolumn{2}{c|}{Damen et al.~\cite{Damen2021RESCALING}}    & 42.1	& 21.5 & 14.8 \\ 
\midrule
\multicolumn{2}{c|}{AudioSlowFast~\cite{kazakos2021slow}}    & 46.5	& 22.8	& 15.4 \\ 
\midrule
\multicolumn{2}{c|}{DATAR (deformable+adaptor)}    & \textbf{48.7}	& \textbf{22.9}	& \textbf{16.3} \\  
\bottomrule
\end{tabular}}
\caption{Comparison with state-of-the-arts on Epic-100.}  
\label{table:3}
\end{table}

\begin{table}[t!]
\small
\centering
\scalebox{1.0}{
\begin{tabular}{c|ccc}
\toprule
{\textbf{Model}}               & \textbf{Top-1} & \textbf{Top-5}  \\ 
\midrule
{ without deformable}    & 69.4	& 86.6	 \\ 
\midrule
{ with deformable}    & 70.5	& 89.9	 \\ 
\midrule

\begin{tabular}{@{}c@{}} deformable + $\mathcal{N}$(0, 0.005) \\ \end{tabular}    & 70.2	& 90.1	 \\ \midrule 
\begin{tabular}{@{}c@{}} deformable + $\mathcal{L}$(0, 0.005) \\ \end{tabular}    & 70.5	& 89.6	 \\   
\midrule

{deform.+adaptor: $\lambda=0.2$}    & 70.5	& 90.2 \\ 
\midrule
{ deform.+adaptor: $\lambda=0.005$}    & \textbf{71.5}	& \textbf{90.5}	 \\ 
\bottomrule
\end{tabular}}
\caption{Ablation study on Kinetics-Sounds.}  
\label{table:4}
\end{table}



\section{Conclusion}
We propose a novel deformable audio transformer for audio event classification, DATAR, which consists of deformable audio attention and input adaptor. Moreover, a learnable input adaptor is introduced to alleviate the issue of over-simplifying the input feature of deformable attention. Extensive results demonstrate the effectiveness of the proposed approach.
\bibliographystyle{IEEEbib}
\small{
\bibliography{IEEEbib}}

\end{document}